\documentclass[twocolumn]{revtex4}
\usepackage{graphicx}
\usepackage{amsmath}

\def\eq#1{Eq.~(\ref{#1})}

\begin{document}

\title{L-selectin mediated leukocyte tethering in shear flow is controled by
multiple contacts and cytoskeletal anchorage facilitating fast rebinding events}
\author{Ulrich S. Schwarz}
\affiliation{Theory Division, Max Planck Institute of Colloids and Interfaces, 14424 Potsdam, Germany}
\author{Ronen Alon}
\affiliation{Department of Immunology, Weizmann Institute of Science, Rehovot, 76100 Israel}

\begin{abstract}
L-selectin mediated tethers result in leukocyte rolling only above a
threshold in shear. Here we present biophysical modeling based on
recently published data from flow chamber experiments (Dwir et al.,
J. Cell Biol. 163: 649-659, 2003) which supports the
interpretation that L-selectin mediated tethers below the shear
threshold correspond to single L-selectin carbohydrate bonds
dissociating on the time scale of milliseconds, whereas L-selectin
mediated tethers above the shear threshold are stabilized by multiple
bonds and fast rebinding of broken bonds, resulting in tether
lifetimes on the timescale of $10^{-1}$ seconds. Our calculations for
cluster dissociation suggest that the single molecule rebinding rate
is of the order of $10^4$ Hz. A similar estimate results if increased
tether dissociation for tail-truncated L-selectin mutants above the
shear threshold is modeled as diffusive escape of single receptors
from the rebinding region due to increased mobility. Using computer
simulations, we show that our model yields first order dissociation
kinetics and exponential dependence of tether dissociation rates on
shear stress. Our results suggest that multiple contacts, cytoskeletal
anchorage of L-selectin and local rebinding of ligand play important
roles in L-selectin tether stabilization and progression of tethers
into persistent rolling on endothelial surfaces.
\end{abstract}

\maketitle

\section*{Introduction}

Leukocyte traffiking plays a central role in the immune response of
vertebrates.  Leukocytes constantly circulate in the cardiovascular
system and enter into tissue and lymph through a multi-step process
involving rolling on the endothelium, activation by chemokines,
arrest, and transendothelial migration \cite{c:spri94}. A key molecule
in this process is L-selectin, a leukocyte-expressed adhesion receptor
which is localized to tips of microvilli and binds to glycosylated
ligands on the endothelium.  Its properties are optimized for initial
capture and rolling under physiological shear
\cite{c:alon95,c:alon97}, as confirmed by recent experimental data and
computer simulations \cite{c:chen99,c:chan00}. In contrast to
tethering through other receptor systems like P-selectin, E-selectin
or integrins, appreciable tethering through L-selectin and subsequent
rolling only occurs above a threshold in shear \cite{c:fing96}, even
in cell-free systems \cite{c:alon98,c:gree00}.  Downregulation by low
shear is unique for L-selectin tethers and might be necessary because
L-selectin ligands are constitutively expressed on circulating
leukocytes, platelets and subsets of blood vessels \cite{c:kans96}.

The dissociation rate of single molecular bonds is expected to depend
exponentially on an externally applied steady force (\textit{Bell
equation}) \cite{c:bell78}. Quantitative analysis with regular video
camera (time resolution of 30 ms) of L-selectin tether kinetics in
flow chambers above the shear threshold resulted in first-order
dissociation kinetics, with a force dependence which could be fit well
to the Bell equation, resulting in a force-free dissociation constant
of $6.6$ Hz \cite{c:alon95,c:alon97,c:chen99,c:chen01}. These findings
have been interpreted as signatures of single L-selectin carbohydrate
bonds. However, recent experimental evidence suggests that L-selectin
tether stabilization involves multiple bonds and local rebinding
events.  Evans and coworkers used the bio\-mem\-brane force probe to
measure unbinding rates for single L-selectin bonds as a function of
loading force \cite{c:evan01b}. Modeling bond rupture as thermally
activated escape over a sequence of transition state barriers
increasingly lowered by rising force \cite{c:evan97}, these
experiments revealed two energy barriers along the unbinding
pathway. The inner barrier corresponds to Ca$^{2+}$-dependent binding
through the lectin domain and explains the high strength of L-selectin
mediated tethers required for cell capture from shear flow. Extracting
barrier properties from the dynamic force spectroscopy data allows to
convert them into a plot of dissociation rate as a function of force.
In this way, results from dynamic force spectroscopy and flow chamber
experiments can be compared in a way which is independent of loading
rate. In detail, Evans and coworkers found a 1000-fold increase in
dissociation rate as force rises from 0 to 200 pN, in marked contrast
to tether dissociation kinetics as measured in flow chamber
experiments, which increases at most 10-fold over this range
\cite{c:alon95,c:alon97}.  Therefore additional stabilization has to
be involved with leukocyte tethers mediated by L-selectin. Dwir and
coworkers used flow chambers to study tethering of leukocytes
transfected with tail-modified mutants of L-selectin \cite{c:dwir01}.
They found that tether dissociation increases with increased tail
truncation, possibly since tail truncation leads to decreased
cytoskeletal anchorage and increased mobility. More recently, Dwir and
coworkers found with high speed video camera (time resolution of 2 ms)
that L-selectin tethers form even below the shear threshold at shear
rate $40$ Hz, albeit with a very fast dissociation rate of $250$ Hz,
undetectable with regular camera \cite{uss:dwir03a}. Thus the shear
threshold results from insufficient tether stabilization at low
shear. Using systematic changes in viscosity (which changes shear
stress, but not shear rate), Dwir and coworkers were able to show that
at the shear threshold, tether lifetime is prolongued by a factor of
$14$ due to shear-mediated cell transport over L-selectin ligand. They
suggested that sufficient transport might be needed for formation of
additional bonds.  With more than one bond being present, rebinding
then could provide the tether stabilization observed experimentally.

In this paper, we present a theoretical model for the interplay
between bond rupture, L-selectin mobility and ligand rebinding within
small clusters of L-selectin bonds, which interprets the recent
experimental results in a consistent and quantitative way.
Traditionally, tether dissociation at low ligand density has been
interpreted as single molecule rupture due to the observed first order
dissociation kinetics and a shear dependence which can be fit well to
the Bell equation. Here we demonstrate that the same features result
for small clusters of multiple bonds with fast rebinding. Our results
suggest that the shear threshold corresponds to the formation of
multiple contacts and that single L-selectin bonds decay too rapidly
as to provide functional leukocyte tethers.

\section*{Experiments}

\begin{figure}
\begin{center}
\includegraphics[width=\columnwidth]{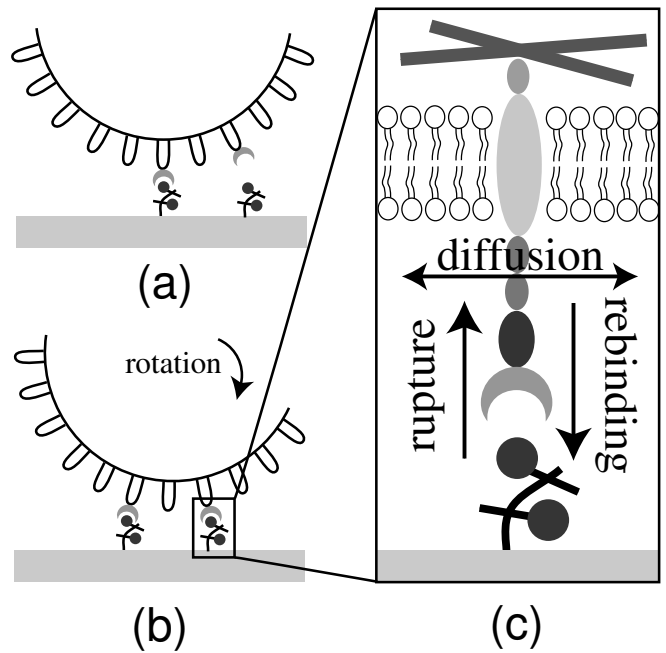}
\end{center}
\caption{Tether dissociation rate $k_{off}$ determined from kinetic analysis
of flow chamber experiments plotted as function of shear rate $\dot
\gamma$ \cite{uss:dwir03a}.  Solid line with circles: wildtype. Dashed line with
diamonds: tail-deleted mutant. Dotted line with squares: wildtype with
6 percent of Ficoll, which changes viscosity and thus
shear stress (but not shear rate) by a factor of 2.6. These data suggest
that the shear threshold is a transport-related rather than a
force-related issue, and that the shear threshold is not about ligand
recognition, but about tether stabilization.}
\label{fig:expkoff}
\end{figure} 

Our experimental procedures have been described 
before elsewhere \cite{c:dwir01,uss:dwir03a}.  Three variants of human
L-selectin were stably expressed in the mouse 300.19 pre B cell
line. Wildtype, tail-truncated mutant and tail-deleted mutant have the
same extracellular domains and differ only in their cytoplasmic
tails. L-selectin mediated tethering was investigated in a parallel
plate flow chamber. The main ligand used was PNAd, the major
L-selectin glycoprotein ligand expressed on endothelium. For
immobilization in the flow chamber, the ligand was diluted in such a
way that no rolling was supported at shear rates lower than 100 Hz
(dilution 10 ng/ml in the coating solution, which corresponds to an
approximate scaffold density of $100 / \mu$m$^2$). Single tethers were
monitored with video microscopy at 2 ms resolution and the
microkinetics were analyzed by single cell tracking as described
previously. The logarithm of the number of cells which pause longer
than time $t$ is plotted as a function of $t$ and usually gives a
straight line indicative of an effectively first order dissociation
process. The slope is the tether dissociation rate $k_{off}$ and is
plotted as a function of shear rate $\dot \gamma$ in
Fig.~1. This plot shows that below the shear threshold of
40 Hz, the dissociation rate is 250 Hz, independent of tail mutations
and viscosity of the medium. Above the shear threshold, the
dissociation rate becomes force-dependent, with a dependence on shear
stress which can be fit well to the Bell equation $k_{off} = k_0 e^{F
/ F_b}$ \cite{c:bell78}. Here $k_0$ is the force-free dissociation
rate and $F_b$ is the bond's internal force scale. The force on an
undeformed 300.19 lymphocyte with radius $R = 6$ $\mu$m follows from
Stokes flow around a sphere close to a wall \cite{c:gold67}. Taking
into account the lever arm geometry provided by the tether holding the
cell at an angle of 50$^{\circ}$, the force acting on the L-selectin
bonds can be calculated to be $F = 180$ pN per dyn/cm$^2$ of shear
stress \cite{c:alon97}. Fitting the Bell equation to the wildtype data
from Fig.~1 gives similar values as obtained in earlier studies
\cite{c:alon95,c:alon97,c:chen99,c:chen01}, namely a force-free
dissociation constant of $6.6$ Hz and an internal force scale $F_b =
200$ pN (corresponding to a reactive compliance of $0.2\
\AA$). At the shear threshold, we find 14-fold and 7-fold reduction in
dissociation rate for wildtype and tail-truncated mutant,
respectively. Adding 6 volume percent of the non-toxic sugar Ficoll
increases viscosity from 1 cP to 2.6 cP. Thus shear stress is
increased by a factor of 2.6, whereas shear rate is unchanged. At the
shear threshold, this increases wildtype dissociation 3-fold, roughly
as expected from the fit to the Bell equation. Most importantly, there
is no shift of the shear threshold as a function of shear rate. This
indicates that the shear threshold results from shear-mediated
transport, rather than from a force-dependent process.

\begin{figure}
\begin{center}
\includegraphics[width=\columnwidth]{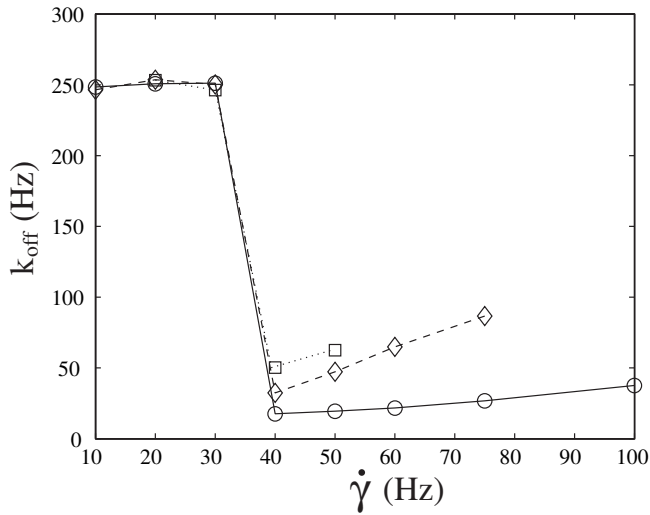}
\end{center}
\caption{A schematic representation of the mechanisms involved
in L-selectin mediated leukocyte tethering to diluted carbohydrate
ligands. (a) Initial binding most likely corresponds to one L-selectin
receptor localized to the tip of one microvillus binding to ligand
presented on a glycoprotein scaffold on the substrate. At low shear,
stabilization through additional bonds is unlikely, because the
distance between scaffolds is larger than the microvilli's tips and
the probability of two microvilli simultaneously hitting two ligands
is very low. (b) At sufficiently high shear, shear-mediated rotation
of the cell over the substrate leads to the establishment of an
additional bond on another microvillus. In contrast to this
two-dimensional cartoon, in practice the two microvilli are expected
to coexist with similar latitude, so that they can share force in a
cooperative way. (c) Close-up to the cell-substrate interface.  The
L-selectin receptor can move laterally in the membrane, with an
effective diffusion constant which depends on cytoskeletal
anchorage. If a receptor has bound to ligand on the substrate, it will
rupture in a stochastic manner, depending on shear-induced loading. If
an additional bond (most likely on the second microvillus) holds the
cell during times of rupture, rebinding can occur at the first
microvillus, thus increasing tether stabilization.}
\label{fig:Lselectin}
\end{figure} 

\section*{Theory}

\textbf{Shear-mediated transport.} At the shear threshold at shear rate 
$\dot \gamma = 40$ Hz (corresponding to shear stress $\tau = \eta \dot
\gamma = 0.4$ dyn/cm$^2$ for viscosity $\eta = 1$ cP) and for small
distance between cell and substrate, a cell with radius $R = 6$ $\mu$m
will translate with hydrodynamic velocity $u = 0.48 R \dot \gamma =
0.12$ $\mu$m/ms and at the same time rotate with frequency $\Omega =
0.26 \dot \gamma = 10.4$ Hz \cite{c:gold67}. Therefore the cell
surface and the substrate surface will move relatively to each other
with an effective velocity $v = u - R \Omega = 0.22 R \dot \gamma =
50$ nm/ms. In average there is no normal force which pushes the cell
onto the substrate, but since it moves in close vicinity to the
substrate, it can explore it with this effective velocity $v$. Thus
there exists a finite probability for a chance encounter between a
L-selectin receptors on the tip of a microvillus and a carbohydrate
ligand on the substrate. Here we focus on the case of diluted ligands,
with a ligand density of $100 / \mu$m$^2$.  Then the average distance
between single ligands is $100$ nm, that is larger than the lateral
extension of the microvilli, which is $80$ nm.  Therefore the first
tethering event is very likely to be a single molecular bond (compare
Fig.~2a). If this first bond has formed, the microvillus
will be pulled straight and the cell will slow down. It will come to a
stop on the distance $x$ of order $\mu$m (e.g. the rest length of a
microvillus is $0.35$ $\mu$m). This takes the typical time $t_s = x/u
= 8$ ms. During this time, the cell can explore an additional distance
of the order of $v t_s = 400$ nm.  The experimental data presented in
Fig.~1 suggest that this is the minimal transport required
to establish a second microvillar contact which is able to contribute
to tether stabilization (compare Fig.~2b).

\textbf{Single bond loading.} If tether duration was much
longer than the time over which the cell comes to a stop, the single
bond dissociation rate $k_{off}$ below the shear threshold should
increase exponentially with shear rate $\dot \gamma$ according to
the Bell equation. However, this assumption is not valid in our case,
because tether duration and slowing down time are both in the
millisecond range. Fig.~3 shows that indeed the Bell
equation (dotted line) does not describe the wildtype data from
Fig.~1 (dashed line with circles). In order to model 
a realistic loading protocol, we assume that the force on the bond
rises linear until time $t_s$ and then plateaus at the constant force
$F$ arising from shear flow.  Note that initial loading rate $r =
F/t_s$ scales quadratically with shear rate $\dot
\gamma$, because $F \sim \dot \gamma$ and $t_s \sim 1 / \dot \gamma$.
The dissociation rate $k_{off}$ for this situation can be calculated
exactly. The result is given in the supplemental material and is
plotted as dash-dotted line in Fig.~3. It is
considerably reduced towards the experimentally observed
plateau. Agreement is expected to increase further if initial loading
is assumed to be sub-linear.  A scaling argument shows the main
mechanism at work.  For the case of pure linear loading, the mean time
to rupture is $T = (F_b / r) \exp(k_0 F_b / r) E(k_0 F_b / r)$, where
$E(x)$ is the exponential integral \cite{c:tees01}. There are two
different scaling regimes for slow and fast loading, which are
separated by the critical loading rate $r_c = k_0 F_b$.  For slow
loading, $r < r_c$, a large argument expansion gives $T \approx
1/k_0$, that is the bond decays by itself before it starts to feel the
effect of force. For fast loading, a small argument expansion gives $T
\approx (F_b / r) \ln (r / k_0 F_b)$, which is also found for the most
frequent time of rupture in this regime \cite{c:evan97}. In our case,
$k_0 = 250$ Hz, $F_b = 200$ pN and $r_c = k_0 F_b = 5 \times 10^4$
pN/s. At the shear threshold, $r = 10^4$ pN/s and we are still in the
regime of slow loading, $r < r_c$. This suggests that tethers below
the shear threshold correspond to single L-selectin carbohydrate bonds
which decay before the effect of force becomes appreciable. This does
not imply that the bonds do not feel any force (after all the cell is
slowed down), but that we are in a regime in which $k_{off}$ as a
function of shear does not change appreciably, as observed
experimentally.

\begin{figure}
\begin{center}
\includegraphics[width=\columnwidth]{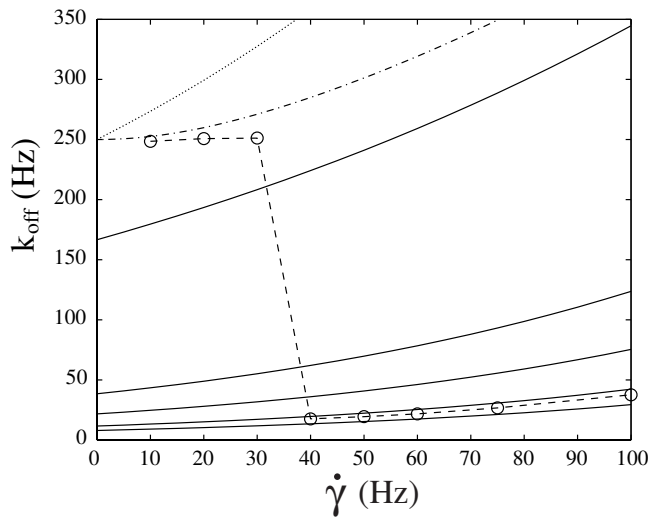}
\end{center}
\caption{Theoretical predictions for tether dissociation 
rate $k_{off}$ as a function of shear rate $\dot \gamma$ compared to
experimentally measured wildtype data from Fig.~1 (dashed
line with circles).  Dotted line: the single bond dissociation rate
with force-free dissociation rate $k_0 = 250$ Hz and constant
instantaneous loading increases exponentially according to the Bell
equation. Dash-dotted line: it is reduced towards the experimentally
observed plateau below the shear threshold at $\dot \gamma = 40$ Hz by
including the effect of finite loading rates.  Solid lines from top to
bottom: cluster dissociation rate for two-bonded tether with rebinding
rate $k_{on} = 0, 10, 20, 40$ and $60\ k_0$. Above the shear
threshold, the two-bonded tether with $k_{on} \approx 10^4$ Hz agrees
well with the experimentally measured data.}
\label{fig:rebindingexact}
\end{figure}  

\textbf{Single bond rebinding.} Single bond rupture is a stochastic
process according to the dissociation rate $k_{off}$ given by the Bell
equation. If ligand and receptor remain in spatial
proximity after rupture, rebinding becomes possible. We define the
single molecule rebinding rate $k_{on}$ to be the rate for bond
formation when receptor and ligand are in close proximity.  If bond
formation was decomposed into transport-determined formation of an
encounter complex and chemical reaction of the two partners, then
$k_{on}$ would correspond to the on-rate for reaction
\cite{c:bell78,r:shou82}.  It has the dimension of 1/s and should not
be confused with two- or three-dimensional association rates, which
have dimensions of m$^2$/s (equivalently m/Ms) and m$^3$/s
(equivalently 1/Ms), respectively. $k_{on}$ should depend mainly on
the extracellular side of the receptor. In the following, it will
therefore be assumed to be the same for wildtype and mutants. There
are two mechanisms which might prevent rebinding within an initially
formed cluster: the single receptor might escape from the rebinding
region due to lateral mobility, or the receptor might be carried away
from the ligand because the cell is carried away by shear
flow. Fig.~2c shows schematically the interplay between
rupture, rebinding and mobility for single L-selectin receptors.

We start with the first case, that is diminished rebinding due to
lateral receptor mobility. Since increased tail truncation decreases
interaction with the cytoskeleton \cite{c:pava95}, lateral mobility
increases from wildtype through tail-truncated to tail-deleted mutant.
For each receptor type, we assume an effective diffusion constant
$D$. The conditional probability for rebinding depends on absolute
time since rupture. We approximate it by the probability that a
particle with two-dimensional diffusion, but without capture is still
within a disc with capture radius $s$ at time $t$, $k_{on}(t) = k_{on}
( 1 - e^{- s^2 / 4 D t})$.  Thus the time scale for the diffusion
correction is set by $s^2 / 4 D$, the time to diffuse the distance of
the capture radius. The diffusion constant for the wildtype can be
estimated to be $10^{-11}$ cm$^2$/s, with the one for the tail-deleted
mutant being at least one order of magnitude larger
\cite{c:chan91,c:kuci96}. A typical value for the capture radius is $s
= 1$ nm. Then the time $t_c = s^2 / 4 D$ to diffuse this distance is
$250$ $\mu$s and $25$ $\mu$s for wildtype and tail-deleted mutant,
respectively. For smaller times, $t < t_c$, $k_{on}$ plateaus at its
initial value.  For larger times, $t > t_c$, it decays rapidly towards
zero. The single molecule behavior is governed by the dimensionless
number $k = k_{on} s^2 / 4 D$, which is the ratio of timescales set by
diffusion and rebinding. Diffusion does not interfere with rebinding
as long as $k > 1$.  Our theory therefore predicts that for wildtype
with diffusion constant $D = 10^{-11}$ cm$^2$/s and capture radius $s
= 1$ nm, $k_{on} > 4 \times 10^3$ Hz. For the tail-deleted mutant,
mobility does interfere with rebinding and we must have $k < 1$.  If
we assume that in this case $D$ is smaller by one order of magnitude,
then $k_{on} < 4 \times 10^4$ Hz.  Thus we can conclude that $k_{on}$
should be of the order of $10^4$ Hz.

\textbf{Tether stabilization through multiple bonds.} We now turn to
the possibility that spatial proximity required for rebinding is
established by multiple contacts. Tethers above the shear threshold
are modeled as clusters of $N$ bonds, which in practice are expected
to be distributed over at least two microvilli.  At any timepoint,
each of the $N$ bonds is either closed or open. The way force is
shared between the closed bonds depends on the details of each tether
realization. However, we expect that only those realizations will
contribute significantly to the long-lived tethers above the shear
threshold in which different bonds share force more or less equally.
This most likely corresponds to two microvilli being bound with
similar latitude in regard to the direction of shear flow. With this
assumption, the force used in the single molecule dissociation rate
has to be overall force divided by the number of closed bonds.  If one
bond ruptures, force is redistributed among the remaining bonds. Open
bonds can rebind with the rebinding rate $k_{on}$.  If rebinding
occurs, force again is redistributed among the closed bonds. In
general, in the absence of diffusion cluster lifetime $T$, but not the
full cluster dissociation probability function can be calculated
exactly \cite{uss:erdm04a}. We first discuss the case without loading
or diffusion, thus focusing on the role of rebinding. As argued in the
supplemental material, for small rebinding rate, $k_{on} < k_0$,
cluster lifetime $T$ scales logarithmically rather than linear with
cluster lifetime $N$. This weak increase in $T$ with $N$ results
because different bonds decay not one after the other, but on the same
time scale. The exact treatment shows that for clusters of $2$, $10$,
$100$, $1000$ and $10,000$ bonds without rebinding, lifetime is
prolongued by $1.5$, $2.9$, $5.2$, $7.5$ and $9.8$, respectively. In
order to achieve 14-fold stabilization as observed experimentally at
the shear threshold, one needs the astronomical number of $6 \times
10^5$ bonds.  In practice, for the case of dilute ligand discussed
here, only very few bonds are likely.  Therefore even in the presence
of multiple bonds, rebinding is essential to provide tether
stabilization.

\begin{figure}
\begin{center}
\includegraphics[width=\columnwidth]{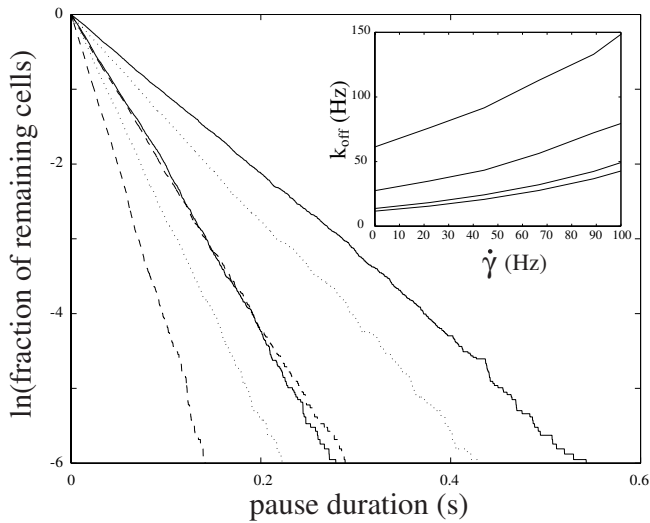}
\end{center}
\caption{Computer simulations show that L-selectin mediated
tethers above the shear threshold yield first order dissociation
kinetics. Solid lines: two-bonded tether with $F = 0$ and $k_{on} =
10^4$ (right) and $0.5 \times 10^4$ Hz (left). Dashed lines: the same
with $F = 100$ pN. Dotted lines: $k_{on} = 10^4$, $F = 0$ and
mobility parameter $k = k_{on} s^2 / 4 D = 1$ (right) and $0.5$
(left), respectively. Inset: L-selectin mediated tethers show
Bell-like shear dependence even in the presence of L-selectin
mobility.  Solid lines from bottom to top: no mobility, $k = 2.5, 1$
and $0.5$.}
\label{fig:simulations}
\end{figure} 

In general, fast rebinding is much more efficient for tether
stabilization than large cluster size. Our calculations predict that
in order to obtain 14-fold stabilization for the cases $N = 2$, $3$
and $4$, one needs $k_{on} = 6 \times 10^3$, $10^3$ and $550$ Hz,
respectively. The value $k_{on} = 6 \times 10^3$ Hz obtained for the
case $N = 2$ is surprisingly close to the estimate $k_{on} = 10^4$ Hz
obtained above via a completely different route, namely the
competition of rebinding and diffusion for a single
molecule. Therefore in the following we restrict ourselves to the
simple case of two bonds being formed above the shear threshold (most
probably by two microvilli). In this case, cluster lifetime can be
calculated to be \cite{uss:erdm04a}:
\begin{equation} \label{eq:rebindingexact}
T = \frac{1}{2 k_0} \left( e^{-F/2 F_b} + 2 e^{-F/F_b} + \frac{k_{on}}{k_0} e^{-3 F/2 F_b} \right)\ .
\end{equation}
A derivation of this result is given in the supplemental material. In
Fig.~3 we use \eq{eq:rebindingexact} to plot the
dissociation rate for the two-bonded tether (identified with the
inverse of cluster lifetime $T$) as a function of shear rate for
different values of rebinding. The shear threshold at
$40$ Hz corresponds to $F = 0.36\ F_b$.  It follows from
\eq{eq:rebindingexact} that for this value of $F$, 14-fold
stabilization in comparision with the force-free single bond lifetime
is achieved for $k_{on} \approx 44\ k_0$.  For $k_0 = 250$ Hz, this
corresponds to a rebinding rate of $k_{on} = 1.1 \times 10^4$ Hz. Thus
again we arrive at the same order of magnitude estimate, $k_{on} =
10^4$ Hz. Fig.~3 shows that with this value for $k_{on}$, agreement
between theory and experimental wild-type data above the shear
threshold is surprisingly good.

\textbf{Relation to BIAcore.} We now discuss how our estimate 
relates to BIAcore data for L-selectin \cite{c:nich98}. In this
experiment, L-selectin was free in solution and GlyCAM-1 immobilized
on the sensor surface, which makes it a monovalent ligand. For the
equilibrium dissociation constant, the authors found $K_D = 105\
\mu$M. This unusually low affinity results from a very large
dissociation rate $k_r$, which they estimated to be $k_r \gtrsim 10$
Hz. The results presented in Fig.~1 seem to suggest that
the real dissociation rate $k_r = 250$ Hz. However, surface anchorage
of both counter-receptors often reduces bond lifetime by up to two
orders of magnitude \cite{c:nguy03}. This has been demonstrated
experimentally for several receptor-ligand systems and might result
from the reduction in free enthalpy of the anchored bond. Thus it
might well be that the dissociation rate $k_0 = 250$ Hz found for
surface anchored bonds might be reduced down to $k_r = 10$ Hz for free
L-selectin binding to surface-bound ligand. Then the association rate
$k_f = 10^5$ 1/Ms. For a capture radius $s = 1$ nm and a
three-dimensional diffusion constant $D = 10^{-6}$ cm$^2$/s, the
diffusive forward rate in solution is $k_+ = 4 \pi D s = 8 \times
10^8$ 1/Ms.  Because $k_+ > k_f$, the receptor-ligand binding in
solution is reaction-limited, as it usually is. As explained above,
$k_{on}$ can be identified with the rate with which an encounter
complex tranforms into the final product \cite{c:bell78,r:shou82}.
Since bond formation is reaction-limited, $k_{on} = k_f K_{+} = 4
\times 10^4$ Hz, where $K_{+} = 3 / 4 \pi s^3$ is the dissociation
constant of diffusion. Thus this estimate agrees well with the
two other estimates derived above.

\section*{Computer simulations}

In order to obtain effective dissociation rates in the presence of
diffusion, one has to use computer simulations. For each parameter set
of interest, we used Monte Carlo simulations to simulate 5,000
realizations according to the rates given above. More details are
described in the supplemental material. In general, our simulations
show that for strong rebinding, that is $k_{on} > k_0$, the effective
dissociation kinetics of small clusters is first order. In Fig.~4,
this is demonstrated for the case $N = 2$. The plot shows the
logarithm of the simulated number of tethers lasting longer than time
$t$ for different parameter values of interest. All curves are linear,
even in the presence of mobility, and the slopes can be identified
with the dissociation rates.  For example, $k_{on} = 10^4$ Hz and $F =
100$ pN yields the same effective first order dissociation rate as
$k_{on} = 0.5 \times 10^4$ Hz and $F = 0$, thus rebinding can rescue
the effect of force. Our simulations also show that cluster
dissocation rate as a function of force fits well to the Bell equation
for $k_{on} > k_0$.  In particular, this holds true in the presence of
L-selectin mobility, as shown in the inset of Fig.~4. For $k_{on} =
10^4$ Hz and without mobility (vanishing diffusion constant), lifetime
at the shear threshold is 12-fold increased compared with single bond
dissociation. With our estimate for wildtype mobility ($k = k_{on} s^2
/ 4 D = 2.5$), 10-fold stabilization takes place.  For tail-deleted
mobility ($k = 0.25$), only 1.5-fold stabilization occurs. This effect
is more dramatic than observed experimentally, where stabilization for
wildtype and tail-deleted mutant are 14-fold and 7-fold,
respectively. In practice, the mobility scenario is certainly more
complicated and is expected to smooth out the threshold effect arising
from our modeling.

\section*{Discussion}

In this paper, we have presented biophysical modeling of L-selectin
tether stabilization in shear flow based on recently published flow
chamber data with high temporal resolution \cite{uss:dwir03a}.  Our
analysis suggests that the 14-fold stabilization observed at the shear
threshold results from formation of multiple contacts and a single
molecule rebinding rate of the order of $k_{on} = 10^4$ Hz, which is
remarkably faster than the force-independent dissociation rate $k_0 =
250$ Hz observed below the shear threshold. Using computer
simulations, we showed that for such strong rebinding, the
experimentally observed first order dissociation and Bell-like shear
force dependence follow from the statistics of small clusters of
bonds. Despite the good quantitative agreement achieved here between
experimental data and our model, it is important to state that it
cannot be expected to predict all details of the experimental
results. In practice, the formation of bonds is a stochastic process
and there will be a statistical mixture of differently sized and
differently loaded clusters, involving different microvilli and
different scaffolds of L-selectin ligands. Cytoskeletal anchorage of
the different ligand-occupied L-selectin molecules might also change
in time in a complex way. Nevertheless, by focusing on the case of two
bonds (possibly on two different microvilli) with shared loading and
mobility-dependent rebinding, we obtained quantitative explanations for
many conflicting observations from flow chamber experiments and
biomembrane force probes, which have not been interpreted in a
consistent way before.

Several explanations have been proposed for the shear threshold effect
before. Chang and Hammer suggested that faster transport leads to
increased probability for receptor ligand encounter
\cite{c:chan99}. Yet the new high resolution data from flow
chamber experiments indicate that below the shear threshold, the issue
is insufficient stabilization rather than insufficient ligand
recognition \cite{uss:dwir03a}.  Chen and Springer suggested that
increased shear helps to overcome a repulsive barrier, possibly
resulting from negative charges on the mucin-like L-selectin ligands
\cite{c:chen99}. However, Dwir and coworkers showed that
small oligopeptide ligands for L-selectin presented on non-mucin
avidin scaffolds exhibit the same shear dependence as their mucin
counterparts \cite{uss:dwir03a}. Evans and coworkers have argued that
increased shear leads to cell flattening and bond formation
\cite{c:evan01b}. However, Dwir and coworkers found that fixation of
PSGL-1 presenting neutrophils does not change the properties of
tethers formed on low density immobilized L-selectin, while they do
destabilize PSGL-1 tethers to immobilized P-selectin (Dwir and Alon,
unpublished data). These data suggest that cell deformation as well as
stretching and bending of microvilli do not play any significant role
in L-selectin tether stabilization. Recently, the unusual molecular
property of catch bonding has been suggested as explanation for the
shear threshold \cite{c:mars03,c:sara04}. However, the data by Dwir
and coworkers suggests that force-related processes do not account for
the shear threshold of L-selectin mediated tethering
\cite{uss:dwir03a}. Our interpretation of the shear threshold as
resulting from multiple bond formation is supported by experimental
evidence that increased ligand density both rescues the diffusion
defect and abolishes the shear threshold \cite{uss:dwir03a}. The
diffusion defect can also be rescued by anchoring of cell-free tail
mutants of L-selectin to surfaces, allowing them to interact with
leukocytes expressing L-selectin ligands \cite{c:dwir01}.

On all ligands tested, the tail-truncated and more so the tail-deleted
L-selectin mutants support considerably shorter tethers, consistent
with a role for anchorage in these local stabilizaton events.  One
possible explanation is that cytoskeletal anchorage prevents uprooting
of L-selectin from the cell. However, uprooting from the plasma
membrane of neutrophils has been shown to take place on the timescale
of seconds \cite{c:shao99}. The tail-truncated L-selectin mutant has
still two charged residues in the tail, which makes it impossible to
extract it from the membrane in milliseconds. Receptor uprooting from
the cytoskeleton only should lead to microvillus extension, which
however is a slow process and has been shown to stabilize the
longer-lived P-selectin mediated tethers rather than L-selectin
mediated tethers \cite{c:yago02}.  Here we postulated another
possibility for cytoskeltal regulation, namely restriction of lateral
mobility. It has been argued before for integrin-mediated adhesion
that increased receptor mobility due to unbinding from the
cytoskeletal is used to upregulate cell adhesion
\cite{c:chan91,c:kuci96}.  Indeed increased receptor mobility is
favorable for contact \emph{formation}, but here we show that it is
unfavorable for contact \emph{maintenance}, since it reduces the
probability for rebinding.

Our analysis suggests that the smallest functional tethers are
mediated by a least two L-selectin bonds, each on a different
microvillus, working cooperatively as one small cluster. Our model
does not explain from which configuration a broken bond rebinds, but
it suggests that this configuration is neither collapsed (otherwise
rebinding, which implies spatial proximity, was not possible) nor
strongly occupied (otherwise diffusive escape was not possible).  We
can only speculate that complete rupture is a multi-stage process, and
that the rebinding discussed here starts from some partially ruptured
state. We also cannot exclude that the rebinding events described here
involve different partners than the dissociated ones, because both
L-selectins and their carbohydrate ligands might be organized in a
dimeric way. Moreover, cytoplasmic anchorage might proceed in multiple
steps, including some weak pre-ligand binding anchorage, which is
strengthened by L-selectin occupancy with ligand. Coupling between
ligand binding and cytoplasmic anchorage is well-known for integrins
\cite{c:hyne02} and might also be at work with selectins. 

The mechanisms discussed in this paper could be effective also with other
vascular counterreceptors specialized to operate under shear flow. As
argued here, the exceptional capacity of L-selectin to promote
functional adhesion in shear flow might not only result from fast
dissociation and high strength under loading, but more so from a fast
rebinding rate. Indeed other vascular adhesion receptors specialized
to capture cells share on-rates similar to that of L-selectin
\cite{c:merw03}. Shear flow may also promote multi-contact formation
for shear-promoted platelet tethering to von Willebrand factor
\cite{c:dogg02}. It may also enhance formation of multivalent
$\alpha_4 \beta_7$ and LFA-1 integrin tethers to their respective
ligands \cite{c:chat01,c:sala02}. The importance of cytoskeletal
anchorage in local rebinding processes of these and related adhesion
receptors has not been experimentally demonstrated to date. However,
the lesson drawn here from the role of L-selectin anchorage in
millisecond tether stabilization may apply to these receptors as well.
Future studies will help confirm this hypothesis. They will also shed
light on the specialized structural features acquired by these
receptors and their ligands through evolution, allowing them to
operate under the versatile conditions of vascular shear flow.

\textbf{Acknowlegdments:} We thank Oren Dwir, Thorsten Erdmann, 
Evan Evans, Stefan Klumpp, Rudolf Merkel, Samuel Safran and Udo
Seifert for helpful discussions. R.A. is the Incumbent of The Tauro
Career Development Chair in Biomedical Research. U.S.S. is supported
by the German Science Foundation through the Emmy Noether Program.

\section*{Supplemental material}

\textbf{Model.} We consider a cluster with a constant number $N$ of 
parallel bonds under constant force $F$. At any time $t$, $i$ bonds
are closed and $N-i$ bonds are open ($0 \le i \le N$). Closed bonds
are assumed to rupture according to the Bell equation \cite{c:bell78}:
\begin{equation} \label{eq:bell}
k_{off} = k_0 e^{F / F_b}\ .
\end{equation}
For convenience, we introduce dimensionless variables: dimensionless
time $\tau = k_0 t$, dimensionless dissociation rate $k_{off}/k_0$ and
dimensionless force $f = F / F_b$. The $i$ closed bonds are assumed to
share force $f$ equally, that is each closed bond is subject to the
force $f/i$. Thus the dimensionless dissociation rate is $e^{f/i}$. As
long as the receptors are held in proximity to the ligands, rebinding
of open bonds can occur. Therefore we assume that single open bonds
rebind with the force independent association rate $k_{on}$. The
dimensionless rebinding rate is defined as $\gamma = k_{on} / k_0$.

The stochastic dynamics of the bond cluster can be described by a
one-step Master equation \cite{uss:erdm04a}
\begin{equation} \label{MasterEquation}
\frac{dp_i}{d\tau} = r_{i+1} p_{i+1} + g_{i-1} p_{i-1} - [ r_i + g_i ] p_i
\end{equation}
where $p_i(\tau)$ is the probability that $i$ closed bonds are present
at time $\tau$. The reverse and forward rates between the different
states $i$ follow from the single molecule rates as
\begin{equation} \label{Rates}
r_i = i e^{f / i}\ , \quad g_i = \gamma (N - i)\ .
\end{equation}
Once the completely dissociated state $i = 0$ is reached, the cell
will be carried away by shear flow and the cluster cannot regain
stability. This corresponds to an absorbing boundary at $i=0$, which
can be implemented by setting $r_0 = g_0 = 0$. Cluster lifetime $T$ is
identified with the mean time to reach the absorbing state $i = 0$.

\textbf{Lifetime of two bonded cluster.} For a cluster with two bonds, 
$N = 2$, cluster lifetime $T$ can be calculated exactly in the
following way. At $\tau = 0$, the cluster starts with the initial
condition $i = 2$. Next it moves to state $i=1$ with probability $1$,
after the mean time $1/r_2$. From there, it rebinds to state $i=2$
with probability $w_R = g_1/(r_1+g_1)$, or dissociates with
probability $w_D = r_1 / (r_1+g_1)$. The mean time for this part is
$1/(r_1+g_1)$. Thus after two steps the system has reached state $i=0$
with probability $w_D$ or returned to state $i=2$ with probability
$w_R$, with $w_D + w_R = 1$. In detail, the probabilities and mean
times for both processes are
\begin{align}
w_D = \frac{e^f}{e^f+\gamma}, \qquad & t_D = \frac{1}{2 e^{f/2}} + \frac{1}{e^f+\gamma}\ , \\
w_R = \frac{\gamma}{e^f+\gamma}, \qquad & t_R = t_D\ .
\end{align}
Different paths to dissociation only differ in the number of
rebinding events $j$ to state $i=2$:
\begin{equation}
w_j = w_D w_R^j, \qquad t_j = t_D + j t_R\ .
\end{equation}
We first check normalization:
\begin{equation}
\sum_{j=0}^{\infty} w_j = w_D \frac{1}{1-w_R} = 1
\end{equation}
and then calculate cluster lifetime:
\begin{align} \label{Ttwobonds}
T & = \sum_{j=0}^{\infty} t_j w_j = t_D + t_R w_D \sum_{j=0}^{\infty} j w_R^j \\
& = t_D + t_R w_D \frac{w_R}{(1-w_R)^2} = \frac{t_D}{1-w_R} \\
\label{eq:T_two_bonds}
& = \frac{1}{2} \left( e^{-f/2} + 2 e^{-f} + \gamma e^{-3f/2} \right)\ .
\end{align}
This formula is given in dimensional form as Eq.~1 in the main text.

\textbf{Cluster size versus rebinding.} For arbitrary cluster size $N$, 
cluster lifetime $T$ can be obtained from the adjoint Master equation
\cite{b:kamp92,uss:erdm04a}. In the case of vanishing force, $f = 0$,
the solution can also be found by using Laplace transforms \cite{uss:erdm04a}:
\begin{equation} \label{eq:VF_lifetime}
T = \frac{1}{(1+\gamma)}\left(\sum_{i = 1}^{N} \binom{N}{i}
\frac{\gamma^i}{i} + \frac{1}{i} \right)\ .
\end{equation}
For $\gamma = 0$, this equation reduces to
\begin{equation} \label{harmonicnumbers}
T = \sum_{i = 1}^{N} \frac{1}{i} = H_N
\end{equation}
where $H_N$ are the harmonic numbers. An expansion for large $N$ gives
\begin{equation}  
H_N = \Gamma + \ln N + \frac{1}{2 N} + O(\frac{1}{N^2})\ .
\end{equation}
Here $\Gamma = 0.577$ is the Euler constant.  This formula is rather
good already for small values of $N$. \eq{harmonicnumbers} is easy to
understand: for $\gamma = 0$, dissociation is simply a sequence of
Poisson decays with mean times $1/r_i = 1/i$. The overall mean time
for dissociation is the sum of the mean times of the subprocesses.  We
conclude that for vanishing rebinding, $T$ grows only weakly
(logarithmically) with $N$ and very large cluster sizes are required
to achieve long lifetimes \cite{c:gold96,c:tees01}. Therefore
rebinding is essential to achieve stabilization for small cluster
sizes.

\textbf{Effect of finite loading rate.} Loading and dissociation 
of single L-selectin bonds occur on the same time scale. As a cell is
captured from shear flow and comes to a stop, force rises from cero
and plateaus at a finite value. We model the initial rise as linear,
with loading rate $r$.  Therefore $f = \mu
\tau$ until time $\tau_0$, followed by constant loading $f = f_0$,
where $\mu = r / k_0 F_b$ is dimensionless loading rate. Since $\mu =
f_0 / \tau_0$, there are only two independent parameters, $\tau_0$ and
$f_0$. The mean lifetime can be calculated in the usual way
\cite{c:evan97,c:tees01}.  We find
\begin{equation} \label{eq:ramp_and_plateau}
T = \frac{e^{\frac{1}{\mu}}}{\mu} \left( E(\frac{1}{\mu}) -
E(\frac{e^{\mu \tau_0}}{\mu}) + \frac{\mu}{e^{f_0}} e^{- \frac{e^{\mu
\tau_0}}{\mu}} \right)
\end{equation}
where $E(x)$ is the exponential integral.  For $\tau_0 \to 0$, we find
the result for constant loading, $T = 1/e^{f_0}$. For $\tau_0 \to
\infty$, we find the result for linear loading, $T = e^{1 / \mu}
E(1/\mu) / \mu$ \cite{c:tees01}. \eq{eq:ramp_and_plateau} is used in
the section on single bond loading and for the plot of the dash-dotted
line in Fig.~3.

\textbf{Simulations.} In the presence of diffusion with 
diffusion constant $D$, the single molecule association rate becomes a
function of the time $t$ which has passed since unbinding. In this
paper, we use the approximation
\begin{equation} \label{eq:kon}
k_{on}(t) = k_{on} \left( 1 - e^{-s^2 / 4 D t} \right)
\end{equation}
where $s$ is the capture radius. Since anaytical solutions are
intractable in this case, the Master equation \eq{MasterEquation} has
to be solved numerically. The standard method to do so are Monte Carlo
simulations. Unfortunately, the Gillespie algorithm for exact
stochastic simulations \cite{c:gill77} cannot be used in this case,
because it does not track the identity of different bonds
\cite{c:firt01}. Therefore we simulate the Master equation 
by discretizing time $\tau$ in small steps $\Delta \tau$.  For
each time step, random numbers are drawn in order to decide how the
system evolves according to the rates defined for the different
processes. In detail, in the time interval $[\tau,\tau+\Delta \tau]$ each
closed bond has the probability $e^{f/i} \Delta \tau$ to rupture, and
each open bond has the probability $\gamma (1-e^{- k / (\gamma \tau)})
\Delta \tau$ to rebind. Here $k = k_{on} s^2 / 4 D$ is the dimensionless
ratio of the timescales set by diffusion and rebinding.  Our model for
stochastic cluster dynamics was implemented in the programming
environment Matlab. A typical run simulates $5,000$ tethers (larger
tether numbers give better statistics for the long time behaviour, but
similar results), each comprising $N$ bonds.  Results from different
runs are bined into histograms for the number of tethers dissociating
in the time interval $[\tau,\tau+\Delta \tau]$. In Fig.~4, we plot the
natural logarithm of the fraction of tethers which last longer than
dimensional time $t$ as a function of $t$, as it is common for the
analysis of experimental data.  The slope of this curve is identified
with the dissociation rate.  Although this procedure involves
numerical integration of the probability distribution for
dissociation, and therefore leads to loss of information, its
smoothing effect is essential to obtain reliable estimates for the
dissociation rate in the presence of noisy data. In the inset of
Fig. 4, the dissociation rates obtained in this way are plotted as
function of shear rate (that is force) and diffusion constant (which
determines the dimensionless parameter $k$).

\end{document}